\documentstyle[twoside,fleqn,epsf,espcrc2]{article}

\title{{\it QCDSP} -- A status report}

\author{{\em The QCDSP Collaboration} --
Dong~Chen,\address{Columbia University, New York, NY}
\thanks{Research supported in part by the U.S. Dept.\ of
Energy.}
\thanks{Current address: CTP-LNS, MIT, Cambridge, MA.}
Ping~Chen,$^{\rm a\;*}$
Norman~H.~Christ,$^{\rm a\;*}$
Robert~G.~Edwards,\address{SCRI, Florida State University,
Tallahassee, FL}$\;^*$
George~R.~Fleming,$^{\rm a\;*}$\thanks{G. Fleming presented poster at Lattice~'97.}
Alan~Gara,\address{Nevis Laboratories, Columbia University,
Irvington, NY}%
\thanks{Research supported in part by the National Science
Foundation.}
Sten~Hansen,\address{Fermi National Accelerator Laboratory,
Batavia, IL}
Chulwoo~Jung,$^{\rm a\;*}$
Adrian~L.~Kaehler,$^{\rm a\;*}$
Anthony~D.~Kennedy,$^{\rm b\;*}$
Gregory~W.~Kilcup,\address{Ohio State University, Columbus, OH }$\;^{*}$
Yubing~Luo,$^{\rm a\;*}$
Catalin~I.~Malureanu,$^{\rm a\;*}$
Robert~D.~Mawhinney,$^{\rm a\;*}$
John~Parsons,$^{\rm c\;\S}$
James~C.~Sexton,\address{Trinity College, Dublin}
ChengZhong~Sui,$^{\rm a\;*}$ and
Pavlos~M.~Vranas$^{\rm a\;*}$} 

\begin{document}

\def\thepage{CU--TP--860}

\begin{abstract}
The {\it QCDSP} machine at Columbia University has grown to 2,048 nodes
achieving a peak speed of 100 Gigaflops. Software for quenched and
Hybrid Monte Carlo (HMC) evolution schemes has been developed
for staggered fermions, with support for Wilson and clover fermions under
development.  We provide an overview of the runtime environment, the current
status of the {\it QCDSP} construction program and preliminary results
not presented elsewhere in these proceedings. 
\end{abstract}

\maketitle

\section{INTRODUCTION}
\label{sec:introduction}

For the past four years \cite{previous} the {\it QCDSP} Collaboration
has developed a simple, scalable parallel computer which exploits the 
latest advances in computer technology.  Using modern computer design
techniques and working closely with hardware manufacturers, we have
constructed several machines at a cost of \$2.7/Mflops in a number of
sizes and configurations.

We begin by summarizing recent progress in upgrading the operating
system and the physics programming environment and discussing the
integration of several QCD-related algorithms.  We present some results
from tests of {\it QCDSP} and the status of our construction program.

\section{SOFTWARE DEVELOPMENT}
\label{sec:sfw_dev}

Software for the {\it QCDSP\/} machine is written in C/C++, with certain
critical routines written in hand-optimized assembly.  

\subsection{The Operating System}
\label{sub:os}

Access to the {\it QCDSP\/} machine is through a SCSI connection to a SUN
workstation.  Users can use an X Windows interface ({\tt qX}) or a standard
UNIX C shell ({\tt qcsh}) with additional commands to control {\it QCDSP}.  
The host environment allows read/write/execute access to all nodes in the
machine, either singly or in groups.

As the machine is booted, the host determines the topology of the SCSI tree
connecting motherboards.  During this process, standard tests are run and a
run-time kernel is loaded and started on the node 0 of each motherboard.  
After probing for more motherboards terminates, all daughterboards in the
machine are booted.  They are checked with standard tests and, if they pass,
their run kernels are loaded.  Finally, global interrupts to all processors
and the 4-dimensional nearest-neighbor communications network are tested.

A user can now issue commands to the {\it QCDSP\/} machine from the host.
Some commands are
\begin{list}{}{
	\settowidth{\labelwidth}{$\tt qset\_nodes \;  3 \; 55$}
	\setlength{\leftmargin}{0in}
	\addtolength{\leftmargin}{\labelwidth}
	\addtolength{\leftmargin}{\labelsep}
}
\item[${\tt qset\_nodes \; all}$] select all nodes
\item[{$\tt qrun \; myprog$}] load and execute a program on the entire machine
\item[{$\tt qset\_nodes \;  3 \; 55$}] select motherboard 3, daughterboard 55
\item[{$\tt qprintf$}] read mb 3, db 5 print buffer to the screen
\item[{$\tt qset\_nodes \; all$}] select all nodes
\item[{$\tt qread \; \ldots$}] read data from all nodes into a file or to the
screen.
\end{list}
The run-time environment on each node is evolving.  Communications libraries,
a disk system and more hardware monitoring are immediate goals.

\subsection{Physics Environment}
\label{sub:phys_env}

We are currently writing the physics environment in C++.  The class
structure accommodates different types of fermions ({\it e.g.} staggered,
Wilson, clover, \dots) and pure gauge actions ({\it e.g.} Wilson plaquette,
Symanzik improved actions, \dots).  The constructor/destructor mechanism
of C++ is used to control initialization and memory allocation. This
mechanism ``hides'' from the user these lower level tasks and at the same
time it guarantees their proper handling.  The virtual class mechanism of
C++ is used to avoid code duplication.  For example the same 
Hybrid Monte Carlo evolution code is used independently of the type of
fermion or pure gauge action.

This code fragment will perform one HMC $\Phi$
trajectory using the Wilson pure gauge action and staggered fermions
followed by a measurement
with conjugate gradient (CG) inversion using Wilson fermions:
\vfill
\small
\begin{verbatim}
main() {
  {
    GwilsonFstag lat;
      // Wilson gauge staggered fermion obj
    AlgHmcPhi hmc(lat, hmc_phi_arg);
      // HMC Phi algorithm object
    hmc.run(1);        // Run the algorithm
  }
  { 
    GwilsonFwilson lat;
      // Wilson gauge Wilson fermion obj 
    AlgCg cg(lat, cg_arg);
      // Conjugate Gradient algorithm obj 
    cg.run();          // Run the algorithm 
  }
} 
\end{verbatim}
\normalsize

There are three evolution algorithms available to the user in the
physics environment:  pure gauge heat bath, staggered HMC $\Phi$ and
HMD $R$ algorithms.  Highly optimized CG inverters for Wilson and
staggered fermions are also available.  Currently, these algorithms run
with 15\%-30\% efficiency, depending on the size of lattice volume per
node.  A clover CG inverter is under development, as are Wilson and
clover HMD fermion force terms.

\def\thepage{\arabic{page}}

\section{TESTING {\it QCDSP}}
\label{sec:testing_qcdsp}

We are conducting a variety of tests using {\it QCDSP}, many involving
physics to check the hardware and software.  The hardware testing begins
when our custom ASIC is manufactured;  Wilson and staggered fermion
conjugate gradients are run at the manufacturer on each piece of
silicon containing an NGA, before it is packaged.  We have continued
this process, by running motherboards of 64 nodes for many hours and
continuously checking the results.

Figure 1 contains the results for the action in
quenched QCD using a variety of machines (including {\it QCDSP}) and a number
of different software algorithms.  Results from machines at Columbia
(filled symbols) and OSU (open symbols) are shown.  hmc is a
hybrid Monte Carlo algorithm, cmhb is a Cabbibo-Marinari heat bath
and metro is a Metropolis algorithm.  The SUN, T3D and ALPHA
results are in double precision, the rest are in single.  The
distribution of actions is well within errors and indicates correct
functioning of {\it QCDSP}.

\begin{figure}[htb]
\epsfxsize=2.9in
\epsfbox[28 84 565 483]{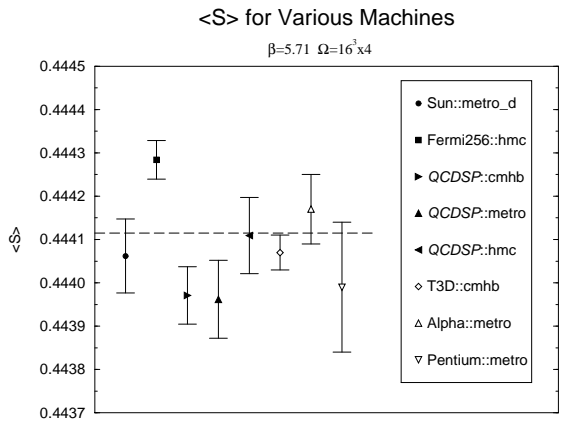}
\vspace{-1cm}
\caption{Test of quenched evolution algorithms}
\end{figure}

We are continuing to study the effects of single precision on the
conjugate gradient algorithm for staggered fermions.  Although {\it
QCDSP} is intrinsically a single precision machine, we have double
precision libraries (where double precision is done in software) that
are supplied by Tartan, Inc., the company which wrote our C++ compiler.
Use of these 64-bit libraries generally costs a factor of 100
in performance.  This makes them not very useful for production
running, but they are extremely useful to check for stability against
finite precision errors.

Our study is as yet incomplete, but our results to date, which
include masses down to $10^{-4}$, show very little difference
between the results for the chiral condensate using single 
and double precision conjugate gradient algorithms

\section{CONSTRUCTION}
\label{sec:construction}

\small 

\begin{table}
\label{tab:construction}
\caption{Construction Program}
\begin{tabular}[ht]{lrc}
{\bf Machine} & {\bf Cost$^\dagger$} & {\bf Completion} \\
\hline
RIKEN/BNL 614 Gflops  & \$1800K & (Dec 97) \\
Columbia 409 Gflops   & \$1800K & (Oct 97) \\
Columbia 102 Gflops   &  \$500K &  Sep 97  \\
SCRI 51 Gflops        &  \$185K &  Aug 97  \\
Ohio State 6.4 Gflops &   \$31K &  Apr 97  \\
Wuppertal 3.2 Gflops  &   \$10K &  Apr 97
\end{tabular}
$^\dagger$ variation in cost/Gflops reflects component volume and cost
at time of purchase.
\end{table}

\normalsize

We have now manufactured nearly 10,000 daughter boards; 56 motherboards;
three single-motherboard enclosures; six eight-motherboard, air-cooled crates; 
and two 16-motherboard, water-cooled cabinets. From these subsystems, we have
constructed a 32 motherboard machine in two cabinets at Columbia 
(see table 1).  We have also assembled a single motherboard
machine (shipped to Wuppertal), a two motherboard machine in a crate 
(shipped to OSU), and a 16 motherboard machine in two crates
(shipped to SCRI).

We are currently assembling the subsystems to expand the 32 motherboard, 102 Gflops,
machine at Columbia to an 8,192 node, 409 Gflops machine in eight cabinets.
The last construction project planned at present is a 12,288 node, 614 Gflops
machine for the RIKEN Brookhaven Research Center, in Brookhaven, New York.
Assembly of the subsystems for this machine is also underway, with final
assembly scheduled for December, 1997.

\end{document}